\documentclass[aps,prd,twocolumn,showpacs,amsmath,amssymb,nofootinbib]
{revtex4-1}
\usepackage{mathptmx}    
\usepackage[T1]{fontenc}
\usepackage{graphicx}
\usepackage{color}
\usepackage{times}
\usepackage{inputenc}
\usepackage{bm}
\usepackage{multirow}
\usepackage{float}
\usepackage{url}
\usepackage{natbib}
\usepackage{mathrsfs}
\usepackage{physics}
\usepackage{comment}
\usepackage[table,xcdraw]{xcolor}
\usepackage{booktabs}
\usepackage{comment}
\usepackage[caption=false]{subfig}
\usepackage[colorlinks=true,citecolor=blue,urlcolor=blue,linkcolor=blue]{hyperref}

\long\def\OFF#1{}

\def\arsinh{\mathop{\text{arsinh}}}

\def\be{\begin{equation}} \def\ee{\end{equation}}
\def\bal#1\eal{\begin{align}#1\end{align}}
\def\ra{\rightarrow}

\def\al{\alpha}
\def\eps{\varepsilon}
\def\la{\Lambda}
\def\l14{\Lambda_{1.4}}
\def\ms{M_\odot}
\def\mmax{M_\text{max}}
\def\mev{\;\text{MeV}}
\def\gev{\;\text{GeV}}
\def\md{\mu}
\def\md{m_\chi}
\def\fd{f_\chi}
\def\Md{M_\chi}
\def\Rd{R_\chi}
\def\gchi{G_\chi}  
\def\cmg{\;\text{cm}^2\!/\text{g}}

\def\olsm#1{\overline{m}_{\nu,\sigma_z}^{(#1)}}
\def\ssn{\sum_{\sigma_z=\pm1} \sum_{\nu=0}^{\nu_\text{max}}}
\def\kfi{k_{F,\nu,\sigma_z}^{(i)}}

\begin{document}

\title{Effects of asymmetric dark matter on a magnetized neutron star:
A two-fluid approach}

\author{Pinku Routaray$^{1}$}
\author{Vishal Parmar$^{2}$}
\author{H. C. Das$^{3}$}
\author{Bharat Kumar$^{1}$}
\author{G. F. Burgio$^{3}$}
\author{H.-J. Schulze$^{3}$}

\affiliation{
$^{1}$ Department of Physics and Astronomy, NIT Rourkela, 769008, India}
\affiliation{
$^{2}$ INFN Sezione di Pisa, Largo B.~Pontecorvo 3, 56127 Pisa, Italy}
\affiliation{
$^{3}$ INFN Sezione di Catania, Dipartimento di Fisica,
Universit\'a di Catania, Via Santa Sofia 64, 95123 Catania, Italy}

\date{\today}

\begin{abstract}
We study the interaction between dark matter (DM)
and highly magnetized neutron stars (NSs),
focusing on how DM particle mass, mass fraction,
and magnetic field (MF) strength affect NS structure and stability.
We consider self-interacting, nonannihilating, asymmetric fermionic DM
that couples to NSs only through gravitational interaction.
Using the Quantum Monte Carlo Relativistic Mean Field (QMC-RMF4) model
with density-dependent magnetic fields,
we investigate the magnetized equation of state
and examine the accumulation of DM under various conditions.
Our results show that as the DM fraction increases,
the maximum gravitational mass of the NS decreases,
especially for heavier DM particles,
while lighter DM particles can induce a transition
from a dark core to a halo structure,
increasing the maximum mass.
Strong MFs soften the equation of state
and reduce the dark mass a NS core can retain
before transitioning to a halo.
By comparing our results with observations from Neutro Star Interior Composition Explorer and GW170817, we identify the possible range of DM parameters for these objects. We find that the magnetic field slightly changes these limits, mainly affecting the maximum NS mass and tidal deformability.
These findings provide key insights into how DM and MF jointly shape
the mass-radius relation
and the stability of DM-admixed magnetized NSs.
\end{abstract}

\maketitle

\section{Introduction}

\begin{table*}[t]
\caption{
Saturation properties and NS observables predicted by the QMC-RMF4 equation of state (EOS) model:
density $\rho_0$,
binding energy per nucleon $E_0$,
compressibility $K_0$,
symmetry energy $S_0$,
its derivative $L_0$,
maximum NS mass $\mmax$,
radii of $1.4\,\ms$ and $2.08\,\ms$ NSs,
and tidal polarizability $\l14$.
Observational ranges are listed for comparison.}
\def\myc#1{\multicolumn{1}{c}{$#1$}}
\renewcommand{\arraystretch}{1.1}
\setlength{\tabcolsep}{3pt}
\begin{ruledtabular}
\begin{tabular}{lccccccccc}
 & $\rho_0$ [fm$^{-3}$] & $-E_0$ [MeV] & $K_0$ [MeV] & $S_0$ [MeV] & $L_0$ [MeV] & $\mmax$ [$\ms$] &$R_{2.08}$ [km] &$R_{1.4}$ [km] & $\la_{1.4}$ \\
\hline
     & 0.162             & 16.1          & 279      & 30.4       & 31.3   & 2.20           & 11.81          & 12.24          & 454 \\
Expt. & $\sim$ 0.14--0.17 & $\sim$ 15--17 & 220--260 & 28.5--34.9 & 30--87 & $>2.35\pm0.17$ & $12.35\pm0.75$ & $12.45\pm0.65$ & 70--580 \\
Ref. & \cite{Margueron18a} & \cite{Margueron18a} & \cite{BALi13,Oertel17}& \cite{BALi13,Oertel17} & \cite{Shlomo06,Piekarewicz10}
     & \cite{Romani22} & \cite{Miller21} & \cite{Miller21} & \cite{Abbott18} \\
\end{tabular}
\end{ruledtabular}
\label{t:eos}
\end{table*}

The majority of matter within the Universe remains obscured in the form of
dark matter (DM) \cite{Bertone05}.
Although numerous discoveries provide compelling evidence
for the existence of DM
\cite{Rubin_Galaxy-rotation_1970,Rubin_Galaxy-rotation_1980,
Clowe_Coma-cluster_2006,Komatsu_Cosmic-Microwave_2011},
none elucidate the particle identity of DM,
which remains an enigma.
Comprehending the characteristics of DM will facilitate
observational astrophysics to identify its nature.
Methods such as direct detection, indirect detection, particle colliders,
and astrophysical probes constitute the means for such observations.
Neutron stars (NSs) stand out as invaluable probes for deciphering
the elusive nature of DM and its scattering cross sections
\cite{Bertone08,Bell21}.
Given their compact and dense composition,
they have become indispensable tools for examining
the particle characteristics of DM.
The incorporation of DM into the constitution of NSs
results in significant modifications of their observables,
including changes in mass-radius profiles, tidal deformability, and luminosity
\cite{Kain21,Rafiei22}.
This phenomenon provides a distinctive indirect avenue
for scrutinizing the properties of DM.

Given the substantial mass of these compact objects,
their presence is predominantly observed in proximity to the Galactic Center
rather than its periphery.
Moreover, the density distribution of DM is higher in these regions.
Consequently, there is a non-negligible probability that DM may be captured
within a NS due to its significant gravitational potential.
The associated accretion rate is contingent upon various factors including
(i) the nature of the DM particle,
(ii) prevailing environmental conditions, and
(iii) the star's internal structure.

Numerous theories have been proposed over many years on the nature of DM
and its impact on the properties of NSs,
considering both gravitational and nongravitational interactions
\cite{Goldman89,Narain06,Kouvaris08,Kouvaris10,Hall10,Kouvaris11,Kouvaris12,
Goldman13,Han14,Grigorious17,Bernal17,Nelson19,Gresham19,
Das20,Ivanytskyi20,Arpan22,Davood22,Leung22,Michael22,Miao22,
Pinku_prd_2023,Rutherford23,Mariani23,Pinku_mnras_2023,Pinku-nitr_2023,Liu23,
Davood24,Shakeri24,Liu24,pinku_ijmpe_2024}.
Among these, asymmetric DM (ADM) interacting gravitationally with NSs
has received significant attention recently,
resulting in several theories on the nature of DM,
both fermionic and bosonic.
In \cite{Nelson19}, the authors considered a trace amount of ADM
captured inside a NS, which self-interacts without annihilation.
They also investigated how DM affects gravitational-wave (GW) emission
and leads to the formation of a DM halo during inspiral,
comparing results for both fermionic and bosonic natures of DM.
Subsequently, in \cite{Ivanytskyi20},
a permissible range for the mass of fermionic ADM and the mass fraction
inside the two massive pulsars PSR~J0348+0432 \cite{Antoniadis13}
and PSR~J0740+6620 \cite{Fonseca21} was obtained.
In \cite{Miao22} a Bayesian analysis was performed
to determine the formation criterion for dark cores/halos
and also investigate the impact of dark halos on the pulsar pulse profile.
Similarly, presuming self-interacting fermionic DM
with dark scalar and vector mediators,
\cite{Arpan22} investigated the impact of DM on NS properties
and carried out a Bayesian analysis
to constrain the DM parameters for a single NS model.
Likewise, other studies assuming self-interacting bosonic ADM
have been conducted to investigate their effects on NS properties
\cite{Davood22,Leung22,Rutherford23,Liu23,Shakeri24,Davood24}.

Nevertheless, the particular effect of DM on highly magnetized NSs
remains a largely unexplored domain.
NSs originate as dense remnants resulting from the explosive collapse
of massive stars during core-collapse supernova events.
Throughout this violent genesis,
the magnetic fields are significantly amplified,
reaching magnitudes ranging from $10^{11}$ to $10^{13}\,$G
\cite{Kaspi10,Kaspi17}.
On rare occasions, magnetars exhibit even more intense fields,
escalating to $10^{14-16}\,$G,
approximately 1000 times stronger than typical pulsars \cite{Makishima14}.
Furthermore, using the scalar virial theorem,
it has been shown that the magnetic field strength in the core can reach
up to $10^{18}\,$G
\cite{chandrasekhar1953problems,Lai91}.
It has also been discussed that the field strength can potentially
be even higher, up to $10^{20}\,$G,
if quark matter exists inside the star
\cite{Ferrer10,Isayev13}.
While the exact origin of such enormous magnetic fields in magnetars
remains a subject of debate,
a widely accepted hypothesis suggests that strong dynamo effects,
driven by the rapid initial spin frequency and differential rotation
during the proto-neutron star phase,
are responsible for amplifying the field to such extreme values
\cite{Lander21,Masada22}.
Additional mechanisms such as magnetorotational instability,
ambipolar diffusion, and Hall drift may also play crucial roles
in sustaining and redistributing the internal magnetic field over time
\cite{Glampedakis10,Lander13}.
These extreme magnetic fields significantly affect the structure and evolution
of NSs, influencing their deformations and, consequently,
leading to substantial GW emissions.

Anomalous x-ray pulsars \cite{Kern02} and
soft gamma repeaters \cite{Hurley11},
subsets of NSs characterized by high magnetic fields,
provide valuable insights.
These pulsars,
distinguished by their intermittent emissions of x-rays and gamma rays,
contribute to the understanding of the complex interplay
between extreme magnetic fields (MFs)
and the observable behavior of NSs \cite{Lattimer21}.
As NSs experience dissipative processes over time,
younger magnetars may exhibit even stronger MFs.
This phenomenon warrants further exploration into the dynamic relationship
between intense MFs and the intrinsic properties of NSs \cite{Gao22}.

In this study, we explore ADM realized through self-interacting fermions,
which interact with highly-magnetized NSs solely via gravitation.
The mass of the DM particle, ranging from MeV to GeV scales,
is treated as a free parameter,
while the repulsive interaction strength
is constrained by observational data on the DM particle cross section.

For the hadronic EOS, we employ the QMC-RMF4 parameter set,
developed within the relativistic mean field (RMF) formalism \cite{Alford22},
and extend it to include density-dependent meson fields
using the methodology described in
\cite{Broderick00,Strickland12,Vishal-magnetised_2023}.
This EOS is capable of producing a NS with a maximum mass of about $2.2\,\ms$.
Considering the gravitational interaction between DM and magnetized NSs,
we examine the impact of DM on macroscopic properties
such as mass, radius, and tidal deformability of magnetized NSs.
A crucial parameter for evaluating these effects is the DM mass fraction,
defined as the ratio of the included DM mass to the total gravitational mass
of the NS.
We investigate the existence of dark cores and dark halos,
accounting for all the free parameters of the DM-admixed magnetized NSs.

This paper is organized as follows.
In Sec.~\ref{s:form} we illustrate the properties of the hadronic EOS
and its extension to the magnetized case;
we also discuss the DM EOS.
In Sec.~\ref{sec:RD} we discuss the results,
and finally in Sec.~\ref{sec:summary} we draw our conclusions.

\section{Formalism}
\label{s:form}

\subsection{Magnetized hadronic EOS}
\label{formulation}

In this work,
we employ a RMF model called ``QMC-RMF4"
that is derived by fitting parameters to the
uniform pure-neutron-matter EOS obtained from chiral effective field theory
\cite{Alford22}.
The unified treatment of the crustal EOS is described in \cite{Parmar22}.
This EOS exhibits stiff behavior,
with a maximum mass of $2.20\,\ms$,
and a canonical radius $R_{1.4}\approx12.3\,$km,
which lies within the limits given by the Neutron Star Interior Composition Explorer (NICER)+XMM data
\cite{Miller21}. 
Also the tidal deformability $\l14$ meets the GW170817 constraint
\cite{Abbott18}.
In Table~\ref{t:eos} we summarize the main properties of the EOS
at saturation density, along with some NS observables.

In the presence of a uniform external MF aligned along the $z$ direction
($\bm{B}=B\hat{\bm{z}}$),
such that $\bm{\nabla}\cdot\bm{B}=0$ \cite{Fang17},
the transverse momenta of charged particles with an electric charge $q$
are quantized into discrete Landau levels \cite{Strickland12}.
The thermodynamic potential $\Omega$ \cite{Ashok01},
which is a function of the chemical potential $\mu$, temperature $T$, and MF $B$,
conforms to canonical relations
$\Omega = -p_\parallel = \eps - \sum_i\rho_i\mu_i$ and
$p_\perp = p_\parallel-MB$.
Here, $\eps$ represents the energy density,
$\rho_i$ denotes the number density of the $i$th particle,
$\mu_i$ the corresponding chemical potential,
$M=-\partial \Omega/\partial B$ represents the system’s magnetization,
and $p_\parallel$ and $p_\perp$ indicate the pressure in the directions
parallel and transverse to the MF, respectively
\cite{Broderick00,Ashok01,Strickland12,Patra20}.

In the present paper,
following the seminal work of \cite{Broderick00},
we compute the magnetized EOS starting from the RMF effective Lagrangian
given in \cite{Muller96,Wang00,Patra02,Kumar20,Das21}.
A detailed description and derivation of the various quantities
required to define the magnetized nuclear matter can be found in
\cite{Broderick00,Ashok01,Strickland12}.
Here, we present the necessary formalism required in the zero-temperature limit.

The energy spectra of neutrons, protons,
and charged leptons (electron and muon) are
\bal
 E_n &= \sqrt{ k^2 + {m_n^*}^2 } + W + R/2 \:,
\\
 E_p &= \sqrt{ k_z^2 + ({\olsm{p}})^2 } + W - R/2 \:,
\\
 E_l &= \sqrt{ k_z^2 + ({\olsm{l}})^2 } \:,
\eal
where $W$ and $R$ are the omega and rho meson mean field, respectively
\cite{Kumar18,Parmar23},
$k_z$ and $\sigma_z=\pm1$ are the momentum and spin
along the direction of the MF,
and $\nu$ is the principal quantum number.
The masses of the charged particles
get modified due to the Landau levels \cite{Broderick00,Strickland12},
\bal \label{e:effmass}
 (\olsm{p})^2 &= {m_p^*}^2
 + 2qB\Big( \nu+\frac12 - \frac12\sigma_z \Big) \:,
\\
 (\olsm{l})^2 &= m_l^2
 - 2qB\Big( \nu+\frac12 + \frac12\sigma_z \Big) \:,
\eal
where $m^*_p$ is the effective mass of the proton.

The partial number and energy densities of the species $i=p,e,\mu$
in presence of the MF are then given by \cite{Broderick00}
\bal
 \rho_i &= \frac{|q|B}{2\pi^2} \ssn \kfi \:,
\label{e:density}
\\
 \eps_i &= \frac{|q|B}{4\pi^2} \ssn
 \Bigg[ E_F^{(i)} \kfi + (\olsm{i})^2
 \arsinh\bigg| \frac{\kfi}{\olsm{i}} \bigg|
 \Bigg] \:.
\label{e:energy}
\eal
In these equations,
the Fermi momentum is defined by
\be\label{e:fermi_momentum}
 \kfi = \sqrt{ {E_F^{(i)}}^2 - {(\olsm{i})}^2 } \:,
\ee
where the Fermi energies $E_F^{(i)}$ are fixed
by the respective chemical potentials,
\bal
 E_F^{(l)} &= \mu_l \:,
\\
 E_F^{(b=p,n)} &= \mu_b - W \pm R/2 \:.
\eal
The largest possible energy label $\nu_\text{max}$
for protons or leptons
is the integer for which the Fermi momentum remains positive, i.e.,
\be\label{e:nmax}
 \nu_\text{max} \leq \frac{E_F^2-{m^*}^2}{2|q|B} \:.
\ee

While the contribution of the neutrons to the pressure is straightforward
\cite{Kumar18,Patra20},
that of the protons can be written in terms of
parallel and perpendicular components
along the local direction of the magnetic field
\cite{Chakrabarty96,Strickland12},
\bal\label{e:ppar}
 p_\parallel &= \frac{|q|B}{4\pi^2} \ssn
 \Bigg[ E_F^{(i)} \kfi - (\olsm{i})^2
 \arsinh\bigg| \frac{\kfi}{\olsm{i}}
 \bigg| \Bigg] \:,
\\ \label{e:pper}
 p_\perp &= \frac{|q|^2B^2}{2\pi^2} \ssn \nu
 \arsinh\bigg| \frac{\kfi}{\olsm{i}} \bigg| \:.
\eal
Consequently, the energy-momentum tensor in the presence of a magnetic field
can be expressed as \cite{Mariani19,Mariani22}
\bal
 T_{\mu\nu} &= T_{\mu\nu}^\text{matter} + T_{\mu\nu}^\text{MF} \:,
\\
 &= \text{diag}\left( \eps+\frac{B^2}{2}, p_\perp+\frac{B^2}{2},
 p_\perp+\frac{B^2}{2}, p_\parallel-\frac{B^2}{2} \right) \:.
\label{e:tmunu}
\eal
In order to be able to employ the standard Tolman-Oppenheimer-Volkoff (TOV)
equations (\ref{e:tovi}--\ref{e:tovnu})
that require an isotropic pressure
(see the extended discussion in \cite{Chatterjee21}),
we average the spatial components of $T_{\mu\nu}$ to obtain an
effective local isotropic pressure.
Adhering to the ``chaotic-magnetic-field'' framework outlined in
\cite{Bednarek03,Lopes15,Flores16,Mariani19},
we express the total average pressure as
\be
 p = \frac{T_{11}+T_{22}+T_{33}}{3}
 = \frac{2p_\perp + p_\parallel}{3} + \frac{B^2}{6} \:.
\label{e:pav}
\ee
Finally, the total energy density $\eps$ and pressure $p$
of the EOS $p(\eps)$ needed in the TOV equations
are obtained for nucleon and lepton contributions as detailed above,
of asymmetric, beta-stable, and charge-neutral matter

subject to the following conditions
relating chemical potentials and partial densities
\cite{Parmar_PRD_2023}.
\be
 \mu_n = \mu_p + \mu_e \:,\quad
 \mu_e = \mu_\mu \:,\quad
 \rho_p = \rho_e + \rho_\mu \:.
\ee
As DM is not interacting directly with ordinary matter,
it is not involved in these equations.
Its EOS is the one of a self-interacting one-component Fermi gas, and
its chemical potential is determined by the gravitational field in the star only.

The anomalous magnetic moment is excluded from our calculations
as it does not significantly affect the EOS \cite{Mariani22}.
Regarding the MF strength profile inside the NS,
we assume the standard parametrization
\cite{Debades97,Rabhi09,Mallick14,Bordbar22}
(but see \cite{Chatterjee21} for a critical discussion)
\be
\label{e:dd_mfield}
 B(\rho) = B_\text{surf} + B_c \Big( 1 - e^{-\beta(\rho/\rho_0)^\gamma} \Big) \:.
\ee
Here, $\rho_0$ is the saturation density,
$B_\text{surf}$ represents the surface MF assumed to be $10^{15}\,$G,
consistent with the observed surface MF of various magnetars
\cite{Tolga11,Casali14}.
$B_c$ pertains to the MF at the core of the star.
The parameters $\beta=0.01$ and $\gamma=3$ are selected
to reproduce the decaying behaviors of the MF \cite{Debades97}.

\subsection{DM EOS}
\label{subsec:DM}

In this study,
the ADM EOS is realized by self-interacting fermions
that do not undergo annihilation
and do not interact with normal matter.
The fermion mass varies from the MeV to GeV scale,
as discussed in \cite{Kain21,Nelson19,Miao22,Liu24},
and the Lagrangian reads
\bal
 {\cal{L}}_\text{DM} &= \Bar{\chi}(i \gamma^\mu D_\mu - \md) \chi
 + \frac12 m_\phi^2\phi_\mu\phi^\mu
 - \frac14 \Omega_{\mu\nu}\Omega^{\mu\nu} \:,
\eal
where $\chi$ and $\phi_\mu$ represent the
fermionic ADM field and vector boson field with masses
$\md$ and $m_\phi$, respectively.
$D_\mu = \partial_\mu+ig_\chi\phi_\mu$
is the covariant derivative,
where $g_\chi$ is the interaction strength of $\chi$ with the $\phi_\mu$ field.
The strength tensor is defined as
$\Omega_{\mu\nu} = \partial_\mu\phi_\nu - \partial_\nu\phi_\mu$.
The corresponding DM energy density and pressure are
\cite{Narain06,Liu24,Barbat24}
\bal
 \eps_\chi &=
 \frac{\md^4}{8\pi^2}
 \Big[ x\sqrt{1+x^2}(2x^2+1) - \arsinh{(x)} \Big]
 + \delta \:,
\\\nonumber
 p_\chi &= \frac{\partial(\eps_\chi/n_\chi)}{\partial n_\chi} n_\chi^2
 = \frac{\partial\eps_\chi}{\partial n_\chi} n_\chi - \eps_\chi
\\
 &= \frac{\md^4}{8\pi^2}
 \Big[ x\sqrt{1+x^2}(2x^2\!/3-1) + \arsinh{(x)} \Big]
 + \delta \:,
\eal
where
\be
 x = \frac{k_\chi}{\md} = \frac{(3\pi^2n_\chi)^{1/3}}{\md}
\ee
is the dimensionless kinetic parameter with the
DM Fermi momentum $k_\chi$ and number density $n_\chi$.
Introducing the dimensionless interaction parameter
$y \equiv g_\chi \md / (\sqrt{2}m_\phi)$,
the self-interaction term is written as
\be
 \delta =
 \left( \frac{yn_\chi}{\md} \right)^2 \:.
\label{e:y}
\ee
Reference ~\cite{Narain06} contains interesting scaling relations
regarding the EOS and mass-radius relations of pure fermionic DM stars.

\OFF
{
We employ in this work the frequently used
\cite{Narain06,Li12,Tulin13,Kouvaris15,Maselli17,Ellis18,Nelson19,
Delpopolo20b,Husain21,Leung22,Collier22,Cassing23}
DM model of fermions with mass $\mu$
self-interacting via a repulsive Yukawa potential
\bal
 V(r) &= \al \frac{e^{-m r}}{r} \:
\eal
with coupling constant $\al$ and mediator mass $m$.
}

Within this model,
$\md$ and $y$ are not independent free parameters,
but constrained
by limits imposed on the DM self-interaction cross section $\sigma_\chi$
through observation of the interaction of galaxies
in different colliding galaxy clusters
\cite{Markevitch04,Kaplinghat16,Sagunski21,Loeb22},
\be
 \sigma_\chi/\md \sim 0.1-10\cmg \:.
\ee
In \cite{Tulin13,Zurek14,Kouvaris15,Maselli17}
it has been shown that the Born approximation
\bal
\OFF
{
 \sigma_\text{Born} &= \frac{4\pi\al^2}{m_\phi^4}\md^2
 = \frac{g_\chi^4 \md^2}{4\pi m_\phi^4}
 = \frac{y^4}{\pi\md^2}
\\
 &\approx 5\times10^{-23} \left(\frac{\al}{0.01}\right)^2
 \left(\frac{10\mev}{m_\phi}\right)^4
 \left(\frac{\md}{10\gev}\right)^2 \text{cm}^2 \:,
\label{e:sigma_chi}
\\
}
 \frac{\sigma_\chi}{\md} &
 = \frac{y^4}{\pi \md^3}
\OFF
{
\\
 &= \frac{10^{12}}{4\pi} \left(\gchi\mev\right)^4
 \frac{\md}{\gev} \frac{1}{\gev^3}
\\
 &= \frac{10^{9}}{4\pi\times4.56} \left(\gchi\mev\right)^4
 \frac{\md}{\gev} \frac{{\rm cm^2}}{\rm g}
\\
 &= 1.75\times10^{7} \left(\gchi\mev\right)^4
 \frac{\md}{\gev} \frac{{\rm cm^2}}{\rm g}
}
\OFF
{
\\
 &\approx 1.78\times10^5 \left(\gchi\mev\right)^4
 \frac{\md}{\gev} \frac{{\rm cm^2}}{\rm g}
\\
 &\approx 1.78\times10^2 \gchi^4 \md \mev^3 \frac{4560}{\gev^3}
\\
 &\approx 1.78\times10^{-7}\times4560 \gchi^4 \md
\\
 &\approx \frac{10^{-2}}{12.32} \gchi^4 \md
}
\eal
is very accurate for $\md\lesssim1\gev$
and in any case remains valid in the limit $y\ra0$ for larger masses.
\OFF
{
With $\gchi=10^{-2}\mev^{-1}$ and for $\md = 0.1-1\gev$,
the values obtained are $\sigma_\chi/\md=0.0175-0.175\cmg$, NOT
consistent with observations \cite{Kaplinghat16,Loeb22}.
Therefore, we consider these values of $\md$ and $\gchi$ in this study.
}
We therefore employ here this approximation,
choosing for simplicity the fixed constraint
\be
 \sigma_\chi/\md = 1\cmg = 4560/\!\gev^3 \:,
\ee
which appears compatible with all current observations.
This implies
\OFF
{
\bal
 \gchi^4 &= \frac{\sigma_\chi}{\md} \frac{4\pi}{\md}
 = \frac{4560}{\gev^3} \frac{4\pi}{\md}
 = \frac{4560\times4\pi}{\gev^4} \frac{\gev}{\md}
\\
 (\gchi\gev)^4 &= {4560\times4\pi} \frac{\gev}{\md}
\\
 \gchi\gev     &= 15.47 \Big(\frac{\gev}{\md}\Big)^{1/4}
 = 15.5 - 27.6
\\
 \gchi\mev     &= 0.0155 \Big(\frac{\gev}{\md}\Big)^{1/4}
 = 0.0155 - 0.0276
(for $\md=1 - 0.1\gev$),
 \gchi\gev &= 15.47 \Big(\frac{\gev}{\md}\Big)^{1/4}
\eal
or equivalently
}
\bal
 y^4 &= \pi \md^3 \sigma_\chi/\md \approx \pi (16.58m_1)^3
\:,\\
 y &\approx 10.94\, m_1^{3/4}
\label{e:ymu}
\eal
with $m_1\equiv\md/1\gev$.
After this, the DM EOS depends only on the one parameter $\md$. 

\begin{figure}[t]
\vskip-1mm
\includegraphics[width=0.5\textwidth]{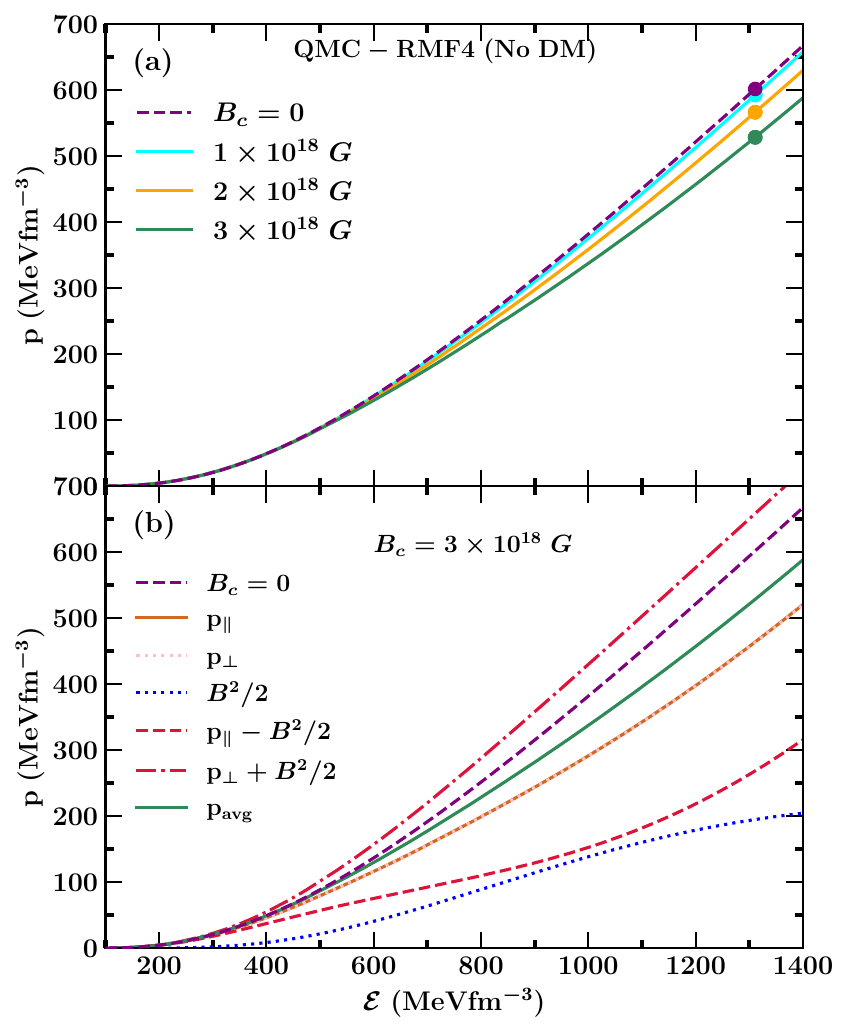}
\vskip-3mm
\caption{
Upper panel (a):
QMC-RMF4 EOS for magnetized NS matter with different MF strengths $B_c$.
The markers indicate the maximum-mass configurations.
Lower panel (b):
Different contributions to the pressure for $B_c=3\times10^{18}\;$G.
}
\label{f:qmc_eos}
\end{figure}

\begin{figure}[t]
\vskip-1mm
\centerline{\includegraphics[width=\columnwidth]{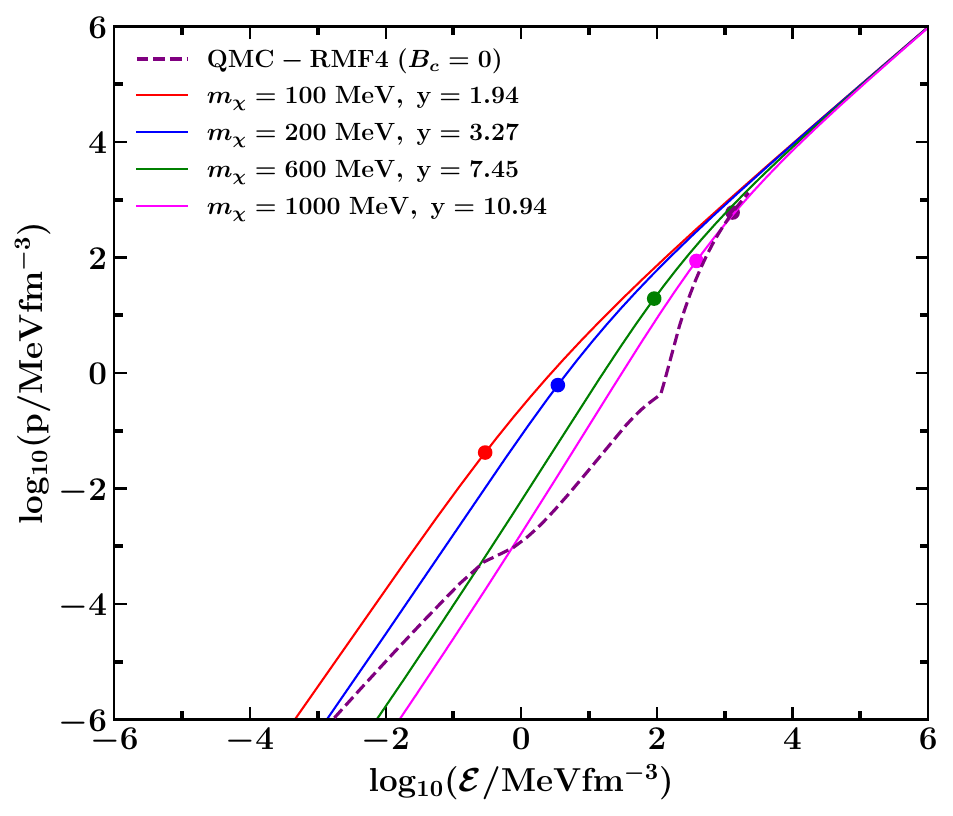}}
\vskip-3mm
\caption{
Pure DM EOS with different masses $\md$
in comparison with the nucleonic QMC-RMF4 EOS.
The DM self-interaction parameter~$y$, Eq.~(\ref{e:ymu}),
is also listed.
The markers indicate the maximum-mass configurations.
}
\label{f:eos_dm}
\end{figure}

\begin{figure*}[t]
\vskip-1mm
\includegraphics[width=1.0\textwidth]{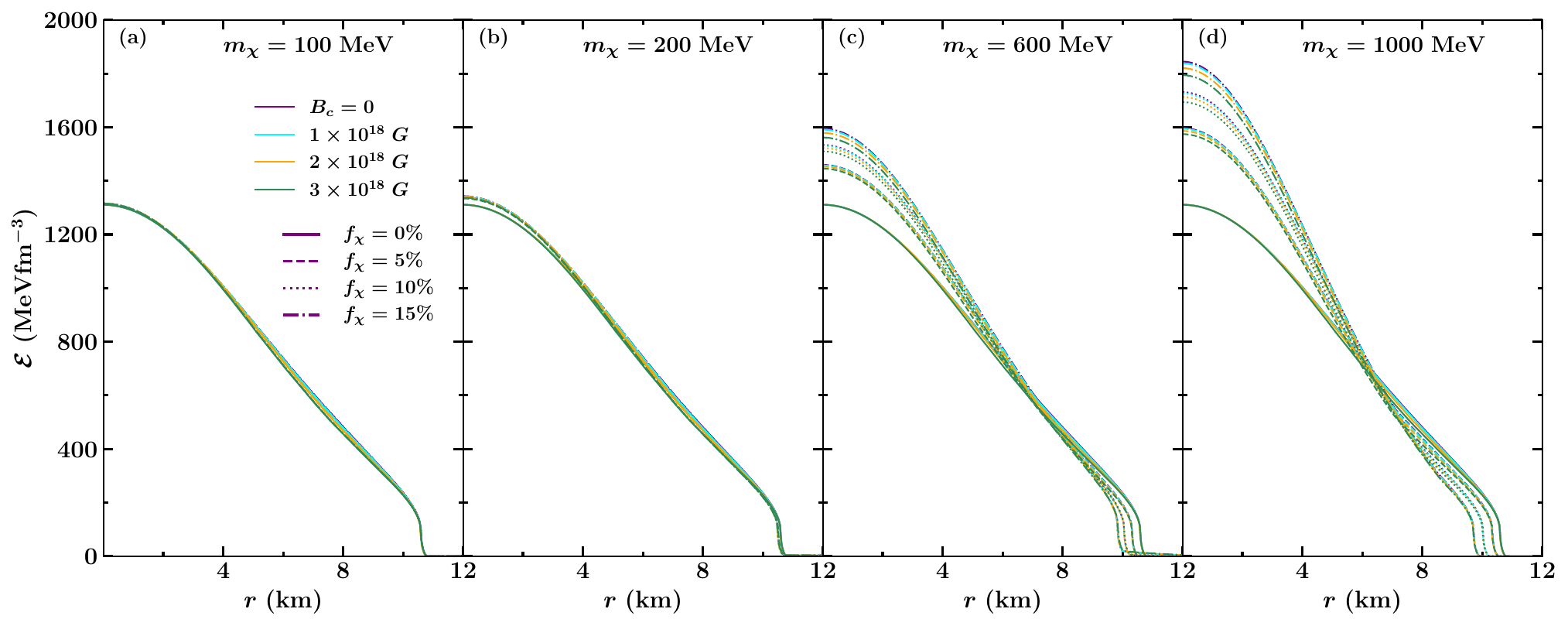}
\vskip-3mm
\caption{
The radial energy-density profiles of magnetized DNSs for the DM EOSs shown in Fig.~\ref{f:eos_dm}. Several choices of DM fraction $f_\chi$ and magnetic field strength $B_c$ are compared at different DM masses in different panels; $m_\chi = 100 \ {\rm MeV}$ (a),  $m_\chi = 200 \ {\rm MeV}$ (b), $m_\chi = 600 \ {\rm MeV}$ (c), $m_\chi = 1000 \ {\rm MeV}$ (d). The curves correspond to the respective maximum-mass configurations.
}
\label{f:eps}
\end{figure*}

\begin{figure*}[t]
\vskip-1mm
\includegraphics[width=1.0\textwidth]{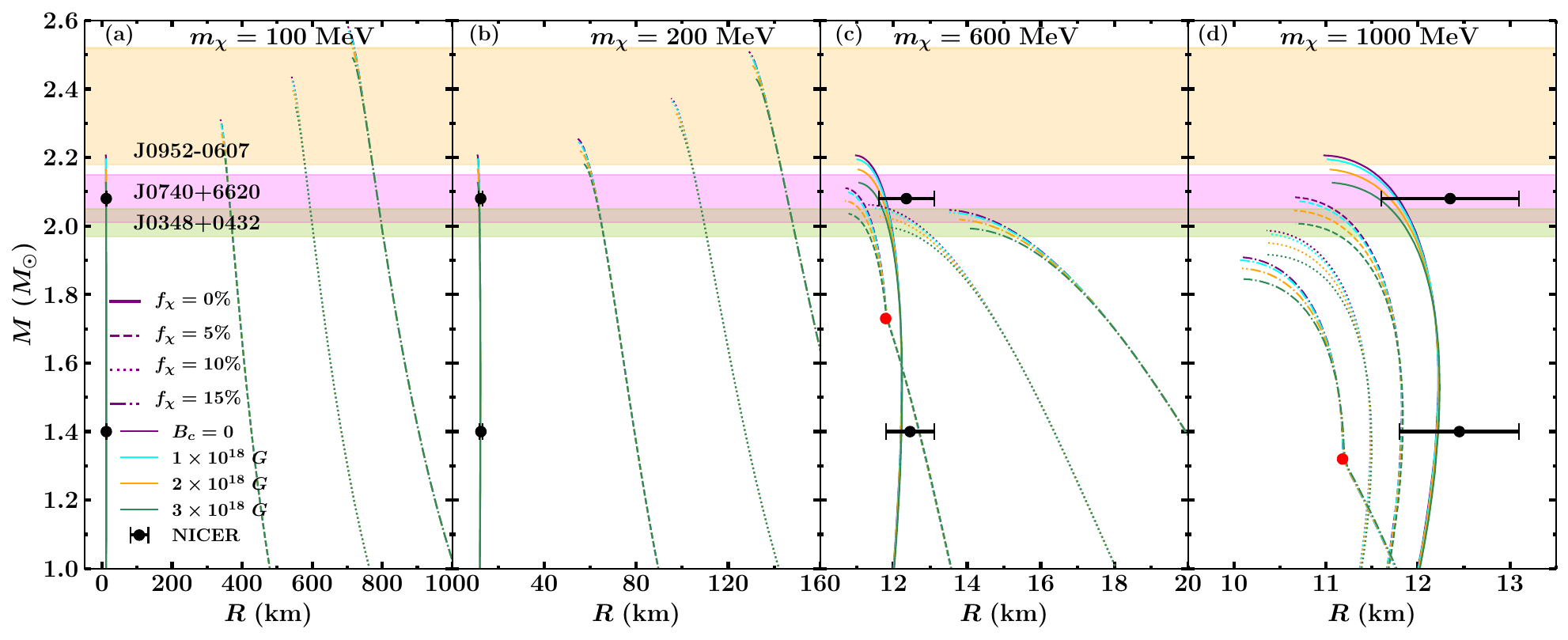}
\vskip-4mm
\caption{
At different $m_\chi$ in different panels like in Fig.~\ref{f:eps}, the mass-radius profiles of magnetized DNS for the DM EOSs and several choices of DM fraction $f_\chi$ and magnetic field strength $B_c$. The radius is $R = {\rm max} (R_N, R_\chi)$, note the different scales. The mass constraints for
PSR~J0952-0607 \cite{Romani22},
PSR~J0740+6620 \cite{Fonseca21}, and
PSR~J0348+0432 \cite{Antoniadis13}
are represented by shaded bars.
The simultaneous $M-R$ constraints from NICER+XMM for
PSR~J0740+6620 \cite{Salmi24}    
and
PSR~J0030+0451 \cite{Vinciguerra24}
are also shown as horizontal error bars.
}
\label{f:mr}
\end{figure*}

\begin{figure}[t]
\vskip-1mm
\centerline{\includegraphics[width=0.5\textwidth]{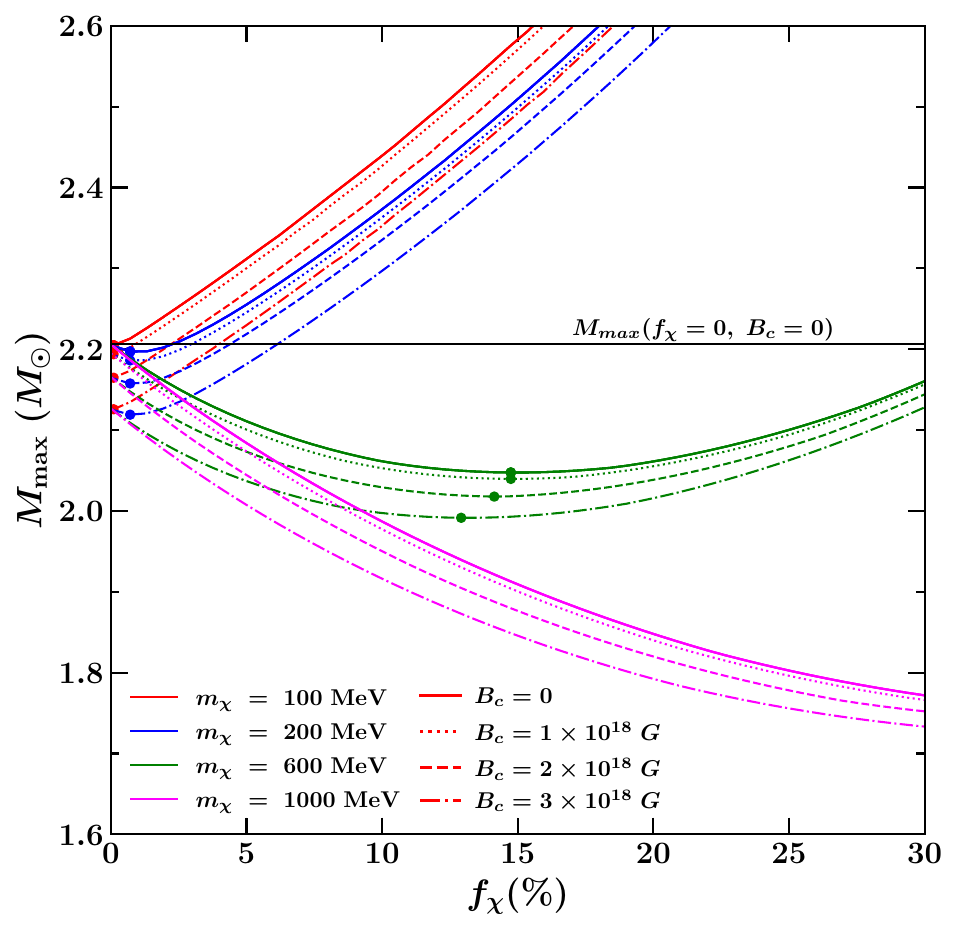}}
\vskip-4mm
\caption{
The maximum gravitational mass as a function of DM fraction
for different DM particle masses and magnetic fields.
The markers indicate the transition from DM core to halo.
}
\label{f:fx-mx}
\end{figure}

\subsection{Hydrostatic equilibrium}
\label{subsec:tov}

The structural properties of the dark matter admixed neutron stars (DNSs) are determined
by solving the two-fluid TOV equations,
assuming that the two non-interacting fluids are in a
static, spherically symmetric, and non-rotating configuration
\cite{Nelson19,Davood22,Leung22,Rutherford23,Liu23,Shakeri24,Davood24}:
\bal
 \frac{dp_i}{dr} &= -(p_i + \eps_i)\frac{d\nu}{dr} \:,
\label{e:tovi}
\\
 \frac{dm_i}{dr}   &= 4\pi r^2 \eps_i \:,
\label{e:tovm}
\\
 \frac{d\nu}{dr} &= \frac{m + 4\pi r^3 p}{r(r - 2m)} \:,
\label{e:tovnu}
\eal
where $i=N,\chi$.
The total pressure, energy density,
and enclosed gravitational mass within a radius $r$ are defined by
$p=p_N+p_\chi$, $\eps=\eps_N+\eps_\chi$, and $m=m_N+m_\chi$.
These coupled differential equations are solved using appropriate
boundary conditions.
At the center of the star, $m_N(0)=m_\chi(0)=0$,
and $p_N(0)=p_c^{(N)}$, $p_\chi(0)=p_c^{(\chi)}$
(which are adjusted iteratively to compute DNSs with
prescribed total $M$ and $\fd$, for example).
The numerical integration continues up to the proper surfaces,
$p_N(R_N)=0$, $p_\chi(R_\chi)=0$, respectively.
The total gravitational mass and its corresponding radius of the system
are defined as
$M = m_N(R_N) + m_\chi (R_\chi)$, and
$R = {\rm max}(R_N, R_\chi)$ respectively.
There are therefore DM-core ($R_\chi<R_N$)
and DM-halo ($R_\chi>R_N$) DNS configurations.

\section{Results and Discussion}
\label{sec:RD}

In this section,
we provide our numerical results for the DNS properties in the presence of a MF.
As mentioned earlier,
the DM has only indirect effects on the properties of the magnetized NS;
therefore, we mainly focus on explaining the results of the combined system
with different scenarios in the following.

\subsection{EOS of magnetized NSs}

In Fig.~\ref{f:qmc_eos}(a)
we present the EOS $p(\eps)$ for nucleonic NSs without DM,
employing the QMC-RMF4 EOS under varying MF strengths
$B_c=1,2,3\times10^{18}\,$G,
comparing with the EOS without a MF ($B_c=0$).
Note that due to the assumption of a density-dependent magnetic field,
Eq.~(\ref{e:dd_mfield}),
$B$ increases with pressure or energy density along the curves,
and this causes a progressive softening of the EOS \cite{Lopes15},
i.e., lower pressures for a given energy density,
resulting in less massive and compact NSs
\cite{Strickland12,Lopes15}.
Thus, the MF's effect on the pressure profile of the NS is considerable,
demonstrating that MFs can induce significant structural changes in NSs.
The markers on the curves indicate the maximum-mass configurations
for each EOS,
anticipating that the maximum mass decreases with stronger MFs.

In the lower panel (b) we show individual contributions to the pressure
at the highest field considered, $B_c=3\times10^{18}\,$G:
The bare matter pressures $p_\perp$ and $p_\parallel$,
Eqs.~(\ref{e:ppar} and \ref{e:pper}),
are practically identical
and strongly reduced compared to the $B=0$ pressure.
The pure field contributions $\pm B^2/2$ to the components of $T_{\mu\nu}$,
Eqs.~(\ref{e:tmunu}),
are of considerable size,
such that also the contribution $B^2/6$
to the average pressure~$p$,
Eqs.~(\ref{e:pav}),
plotted in panel (a),
is sizeable.
However, the effect is not enough to compensate the reduction
of $p_\perp$, $p_\parallel$ relative to the $B=0$ pressure.
Note that the pressures shown in Fig.~\ref{f:qmc_eos}
always contain contributions of neutrons, protons, and leptons.
These results corroborate earlier findings,
which also showed that deviations from spherical symmetry,
even in high-MF scenarios up to $10^{18}\,$G,
remain minimal (less than 1\%)
\cite{Chu15,Patra20,Bordbar22},
thus, supporting the assumption that the structure of highly-magnetized NSs
can still be effectively described assuming spherical symmetry.

\subsection{EOS of ADM}

The DM EOS is depicted in Fig.~\ref{f:eos_dm}
for different DM candidate masses $\md=0.1-1\gev$
and the compatible DM self-interaction parameter $y$, Eq.~(\ref{e:ymu}).
The hadronic QMC-RMF4 EOS is shown for comparison,
exhibiting different domains for core, inner crust, and outer crust.
Markers indicate the maximum-mass configurations of pure dark stars or
standard NSs.
Within the range of interest,
lighter DM masses result in a notably stiffer EOS,
corresponding to larger DNS maximum masses,
as is well known \cite{Kain21,Liu24}.

\subsection{Density profiles}

The interaction between DM and a magnetized NS can result in
either a DM-core or a DM-halo star,
mainly determined by the DM particle mass $\md$
(and the correlated interaction strength),
and the DM mass fraction $\fd = \Md/M$ \cite{Liu23,Liu24}.
We are interested in the effect of a magnetic field on this feature.

In Fig.~\ref{f:eps},
the radial energy-density profiles of magnetized DNSs
are shown for the maximum-mass configuration of each EOS,
varying $\md$, $\fd$, and $B_c$.
The figure illustrates that as the DM particle mass or its fraction increase,
the energy density rises,
corresponding to a more compact star.
As will be better seen in Fig.~\ref{f:mr},
for light DM masses, such as $\md=100,200\mev$,
the DM extends beyond the normal matter radius $R_N$
for all chosen values of $\fd$,
forming a DM-halo star,
whereas for heavier DM masses like $\md=600,1000\mev$,
the DM is entirely confined within the star
regardless of the DM mass fraction $\fd$.
With increasing $\fd$, also the dark radius $\Rd$ increases,
extending further the DM halo for ``small" $\md$,
whereas for ``large" $\md$ the DM remains trapped within the DNS core,
substantially increasing the star's compactness.

An increasing MF also decreases the compactness by a few percent,
but the effect is much weaker than varying the DM fraction in the figure.

\subsection{Mass-radius relations}

The density profiles of magnetized DNSs show clearly that
the model parameters $\md$ and $\fd$
determine the formation of either a DM-halo ($R=\Rd$)
or a DM-core ($R=R_N$) structure
and thus impact significantly the radius and overall structure of the DNS,
whereas the magnetic field appears of minor importance.
This is also seen in the mass-radius profiles shown in Fig.~\ref{f:mr},
for the same conditions as in Fig.~\ref{f:eps}.
As analyzed in more detail in \cite{Liu24},
for small (large) masses $\md \lesssim (\gtrsim) 1\gev$
DM-halo (core) stars are formed,
where both $R$ and $\mmax$ increase (decrease) with increasing
(not too large) $\fd$.
Two typical contrasting cases are shown in the
$\md=100\mev$ and $\md=1000\mev$ panels.

The $\md=600\mev$ panel illustrates the transition between both regimes:
on the $\fd=5\%$ curves one notes a $\Rd=R_N$ configuration
(red marker)
with $M\approx1.73\ms$ and $R\approx11.79\,$km,
where the $R_N$ branch for larger masses deviates onto the $\Rd$ branch
for lower masses.
This transition is continuous across the $M(R)$ curve.
The same occurs on the $\fd=15\%$ curves for $\md=1000\mev$.
Again we refer to \cite{Liu24} for a more extended discussion.

The figure also shows current maximum-mass constraints
from PSR~J0952-0607, PSR~J0740+6620, and PSR~J0348+0432,
as well as NICER+XMM constraints of the radii $R_{1.4}$ and $R_{2.08}$.
The nucleonic RMF EOS is compatible with all of them,
and some admixture of DM can currently not be excluded for the observed objects.

The effect of the magnetic field
(reduction of $\mmax$)
is not more than a few percent for all configurations
and will be analyzed in more detail now.

\begin{figure}[t]
\vskip-4mm
\includegraphics[width=1.0\columnwidth]{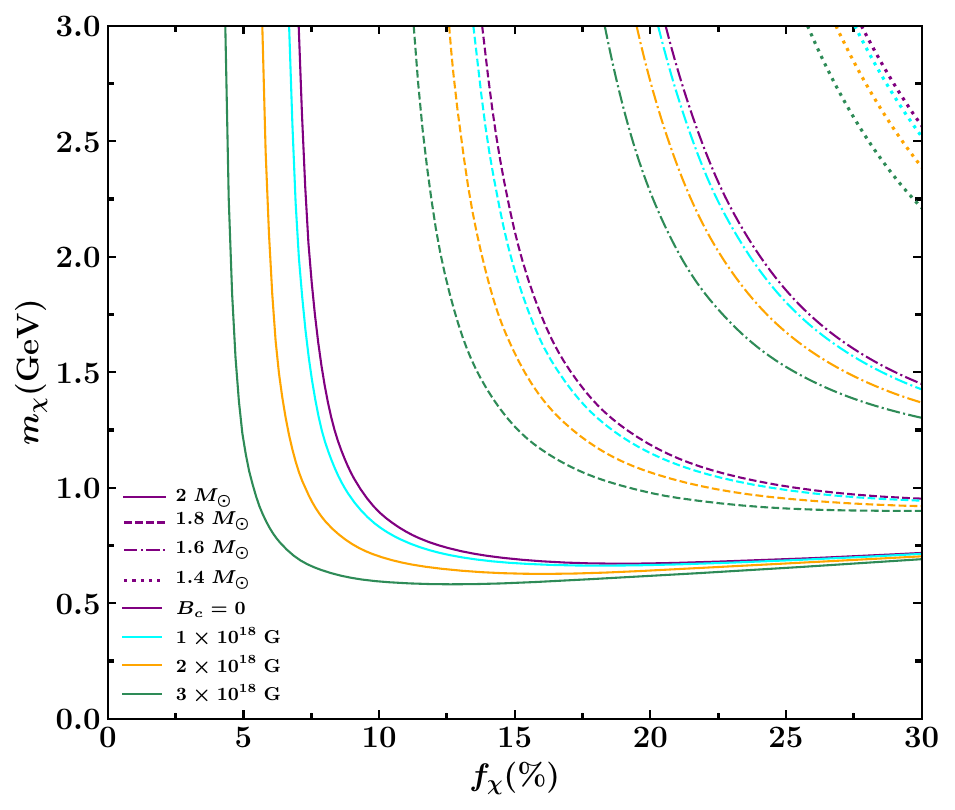}
\vskip-4mm
\caption{
$\mmax$ contours in the $(\md,\fd)$ plane
for different magnetic fields.
}
\label{f:crit_fx}
\end{figure}

\subsection{Maximum mass}
\label{subsec: Max_contour}

In Fig.~\ref{f:fx-mx} we show the maximum DNS gravitational mass
as a function of DM fraction,
for different $\md$ and $B_c$ values.
For each EOS the transition from DM core to halo is indicated by a marker.
In accordance with Fig.~\ref{f:mr},
for the small masses $\md=100,200\mev$ the DM-halo character sets in
at low $\fd<1\%$, increasing $\mmax$,
while for large $\md=600\mev$ the onset occurs at $\fd\approx15\%$,
and for $\md=1000\mev$,
there are only DM-core configurations with lowered $\mmax$ in the plot range.
Again, the effect of the magnetic field is very small,
in particular there is a small reduction of the halo-core transition fraction.

Finally,
Fig.~\ref{f:crit_fx} shows some contours of $\mmax$ in the $(\md,\fd)$ plane
for different magnetic fields.
As the previous figure,
it indicates that for greater $\md$,
a smaller $\fd$ is sufficient to destabilize the DNS.
Since a magnetic field decreases the maximum mass,
it provides some degree of destabilization against DM-induced collapse,
but the effect is again small compared to the variation of $\fd$.

\subsection{Tidal deformability}

\begin{figure}[t]
\vskip-1mm
\centerline{\includegraphics[width=1.0\columnwidth]{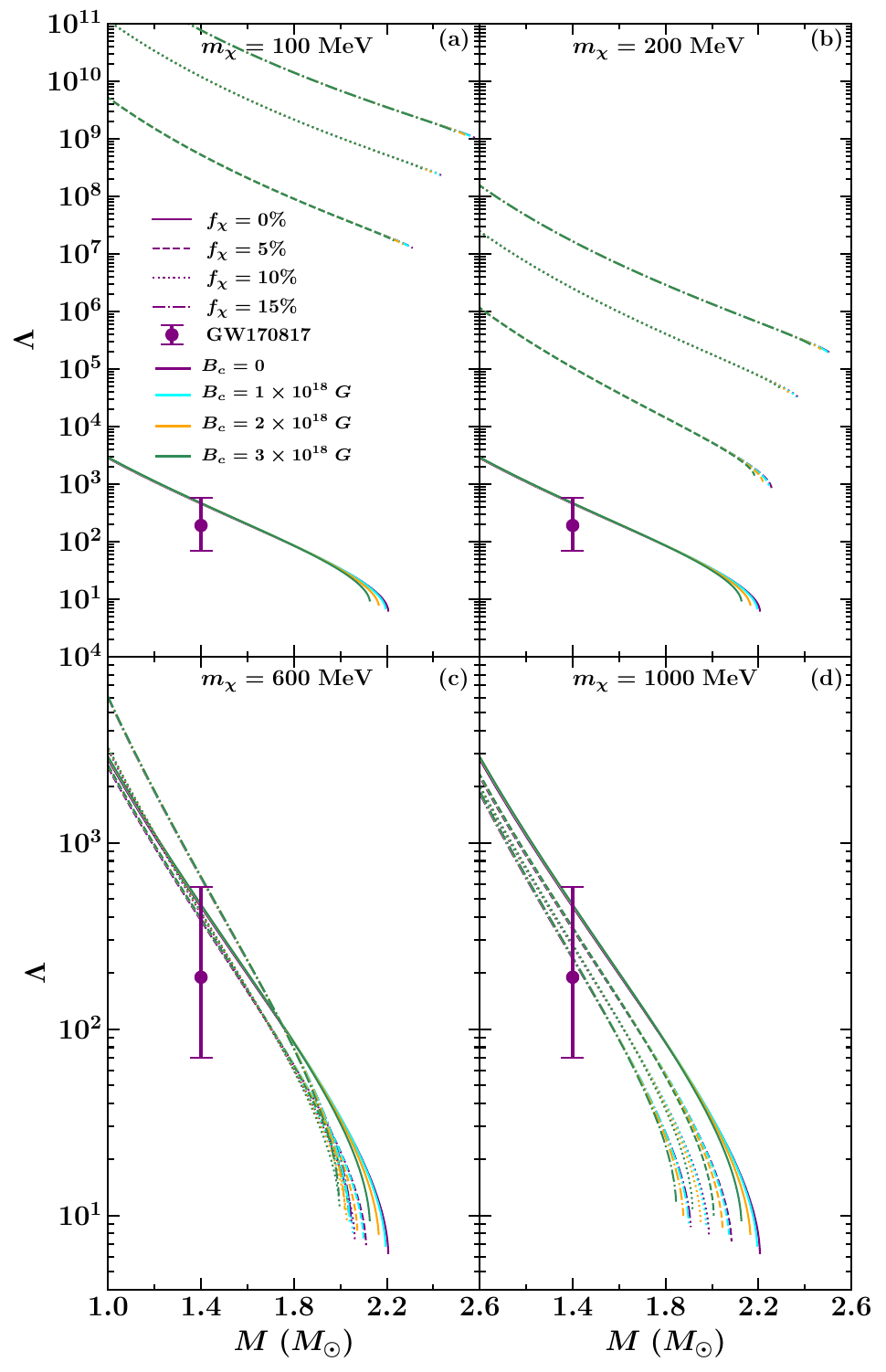}}
\vskip-3mm
\caption{ Similar to Fig.~\ref{f:eps} and Fig.~\ref{f:mr}, at different $m_\chi$ in different panels,
the tidal deformability vs DNS mass
for different values of $\md$, $\fd$, and $B_c$ is displayed.
The GW170817 constraint \cite{Abbott17} is also shown.
}
\label{f:tidal}
\end{figure}

In a binary system, the gravitational interaction with a companion object
(either a NS or a black hole) induces deformation in a NS.
The dimensionless tidal deformability of the system,
which quantifies this deformation,
is represented by
$\Lambda=(2/3)(R/M)^5 k_2$,
where $k_2$ refers to its second Love number \cite{Hinderer08,Kumar17}.
$\Lambda$ depends on the star's mass and radius,
and is modulated by DM and magnetic fields.
Within the two-fluid framework,
it has been examined both with
\cite{Patra20,Rather21,Parmar23}
and without
\cite{Nelson19,Leung22,Michael22,Arpan22,Liu23,Liu24}
the presence of a MF.

In Fig.~\ref{f:tidal}, $\Lambda$ is plotted vs the DNS mass
for different values of $\md$, $\fd$, and $B_c$.
As $\Lambda \sim R^5$ is extremely sensitive to the gravitational radius,
the DM-halo or -core character plays a decisive role:
According to Fig.~\ref{f:mr},
the small-$\md$ DM-halo stars feature very large radii and consequently
enormous values of $\Lambda$,
whereas the large-$\md$ DM-core stars with their reduced radii
exhibit also reduced $\Lambda$ values.

Across all panels it is apparent that
the effect of the MF is very small
and only visible near the $\mmax$ configurations.
A stronger MF results in a more compact NS,
thereby decreasing its deformability.

The pure NS EOS with $\Lambda=454$ fulfills (by construction)
the observational limits imposed by GW170817,
but also here,
the large error bar currently cannot exclude an admixture of DM.

\subsection{Constraining dark matter parameter space}

\begin{figure}[t]
\vskip-1mm
\centerline{\includegraphics[width=1.05\columnwidth]{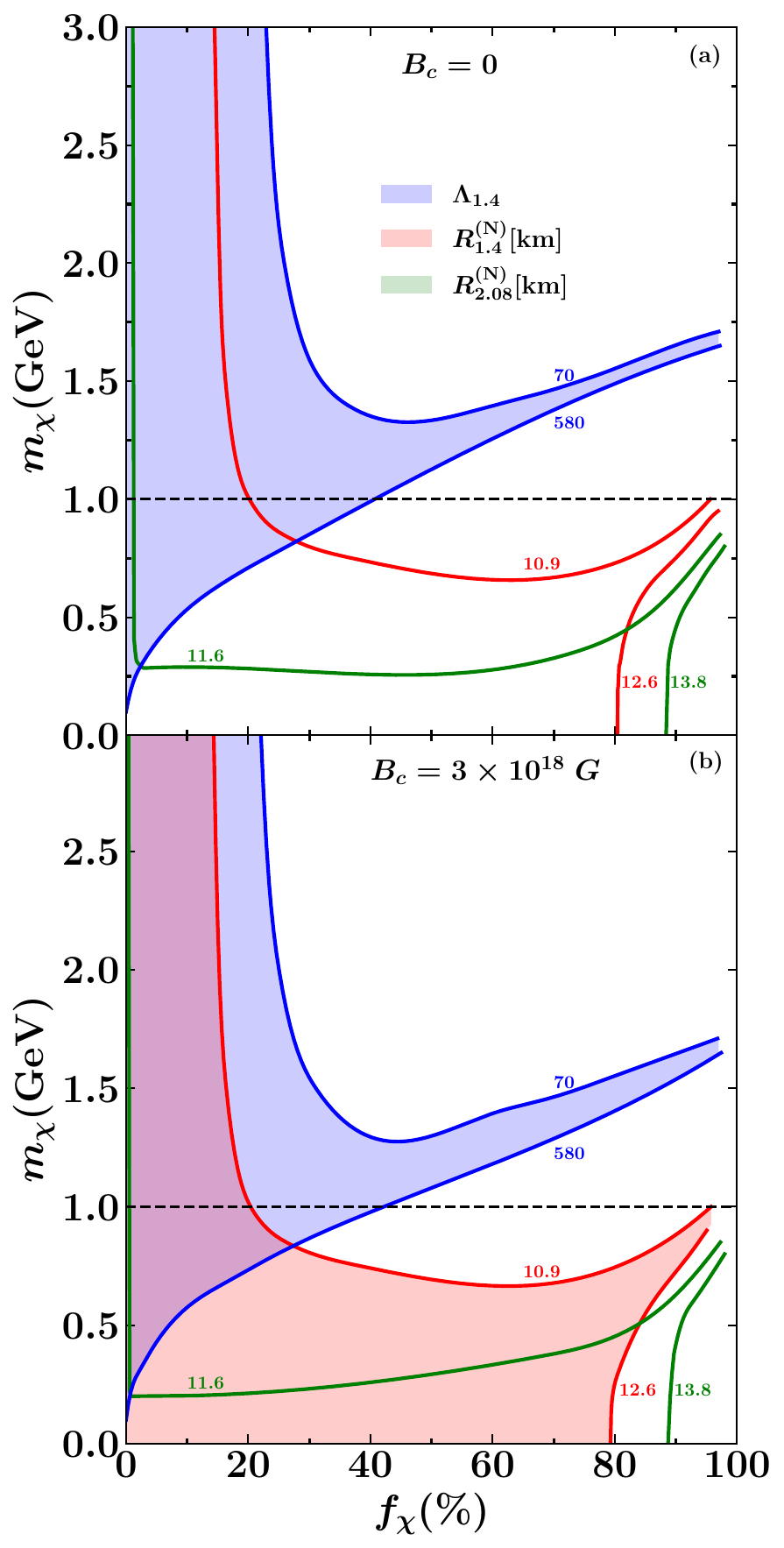}}
\vskip-3mm
\caption{The mass and radius constraints from NICER measurements
for PSR~J0740+6620 \cite{Salmi24} and PSR~J0030+0451 \cite{Vinciguerra24}
and the tidal deformability constraint from GW170817 \cite{Abbott17}
in the $(\md,\fd)$ plane for the QMC-RMF4 model
with (b) and without (a) MF.
}
\label{f:m_chi_f_chi}
\end{figure}

Above, we have discussed observational constraints on NS
mass, radius, and tidal deformability imposed by GW170817 and NICER.
Although the relevant objects are probably just pure NSs,
one may ask the question which kind of DNSs might be accommodated by these
data, if those objects would contain DM.
In particular, optical radius measurements from NICER provide further constraints on the DM parameter space \cite{Vinciguerra24,Salmi24}. These observations suggest that the nuclear core sizes of DM-admixed stars must remain close to those of pure NSs to align with current data.
However, it is important to stress that those and other current observational
data do not impose any constraints on models of DM,
as DNSs with a sizable DM fraction might simply never have been observed so far.

In Fig.~\ref{f:m_chi_f_chi} we nevertheless present the constraints
on the DM parameters $\md$ and $\fd$ imposed in this sense
by some observations, namely

(a) the optical radius
${R_N}_{1.4}=11.7^{+0.9}_{-0.8}\,$km
reported by NICER \cite{Vinciguerra24} 
for PSR~J0030+0451 with $M=1.40\pm0.13\,\ms$;

(b) the optical radius
${R_N}_{2.08}=12.5^{+1.3}_{-0.9}\,$km
reported by NICER \cite{Salmi24} 
for PSR~J0740+6620 with $M=2.08\pm0.07\,\ms$;

(c) the tidal deformability
$70<\l14<580$
from GW170817 with
$M=1.15-1.64\,\ms$
\cite{Abbott17,Abbott18}.

Regarding PSR~J0030+0451,
the red region delimited by the ${R_N}_{1.4}=10.9$ and $12.6\,$km contours
indicates the range of ($\md,\fd$)
permissible for DNSs that would mimic a pure NS with that mass and radius.
Similarly, the green area indicates the allowed parameter range
for PSR~J0740+6620.
In particular, in all cases small enough values of $\fd$ cannot be excluded
for any $\md$,
because in these DM-core stars the nuclear radius stays close to
(slightly below) the one of the pure NS.
Large values of $\fd$ are possible for small $\md<1\gev$,
and correspond to DM-halo stars with a nuclear core that
is not extended too much compared to the pure NS.
However, the normal matter (NM) fraction and density are quite small in those
stars and probably the use of the low-density NS EOS is questionable.
A more detailed discussion can be found in \cite{Liu24}.
Obviously these constraints are specific to the particular objects
and {\em cannot} be combined to a global constraint of DM parameters.

The same holds for the blue region representing the GW170817 constraint
$70<\l14<580$ \cite{Abbott18}.
In particular, values of large $\fd$ $/$ small $\md$ are forbidden,
because they correspond to DM-halo stars with very large values of
$\l14$ \cite{Liu24}.

Finally, it is important to note that
all constraints discussed above
depend most of all on the nucleonic EOS
(and its predictions for $\mmax,R_{1.4},R_{2.08},\l14$
chosen for the analysis.

The figure compares results with $B_c=0$ (a)
and $B_c=3\times10^{18}\,$G (b).
The inclusion of a strong magnetic field slightly modifies (restricts)
the allowed parameter space,
but it is hard to see in the figure,
apart from ${R_N}_{2.08}$,
which is mainly affected by the reduced $\mmax$ of the magnetized star.

\OFF{
The most massive cold NSs observed so far are
PSR~J1614-2230 ($M=1.908\pm0.016\ms$) \cite{Arzoumanian18},
PSR~J0348+0432 ($M=2.01\pm0.04\ms$) \cite{Antoniadis13}, and
PSR~J0740+6620  2.14-0.09+0.10 \cite{Cromartie20}, OLD!
PSR~J0740+6620 ($M=2.08\pm0.07\ms$) \cite{Fonseca21},
PSR~J0952-0607 ($M=2.35\pm0.17\ms$) \cite{Romani22}

PSR~J0740+6620 with measured radius
$R(2.08\pm0.07M_\odot) = 13.7^{+2.6}_{-1.5}\km$ \cite{Riley21}
$R(2.072^{+0.067}_{-0.066}\ms) = 12.39^{+1.30}_{-0.98}\km$ \cite{Miller21}
$R(2.073\pm0.069\ms) = 12.49-0.88+1.28 \cite{Salmi24} 11.6,12.5,13.8
12.76-1.02+1.49 or 12.92-1.13+2.09 \cite{Dittmann24}

We also mention the combined estimates of the mass and radius
of the isolated pulsar PSR~J0030+0451 observed recently by NICER,
$M=1.44^{+0.15}_{-0.14}\ms$ and $R=13.02^{+1.24}_{-1.06}\km$
\cite{Miller19,Riley19}
$M=1.36^{+0.15}_{-0.16}\ms$ and $R=12.71^{+1.14}_{-1.19}\km$ 11.5,12.7,13.8
\cite{Miller21,Riley21}
$M=1.40^{+0.13}_{-0.12}\ms$ and $R=11.71^{+0.88}_{-0.83}\km$ 10.9,11.7,12.6
$M=1.70^{+0.18}_{-0.19}\ms$ and $R=14.44^{+0.88}_{-1.05}\km$
\cite{Vinciguerra24}

and in particular the result of the combined GW170817+NICER analysis
\cite{Abbott17,Abbott18},
$R_{2.08}=12.35\pm0.75\,$km \cite{Miller21}, and
$R_{1.4}=
$12.45\pm0.65$ km \cite{Miller21}, 
$11.94^{+0.76}_{-0.87}$ km \cite{Pang21}, and
$12.33^{+0.76}_{-0.81}$ km or
$12.18^{+0.56}_{-0.79}$ km \cite{Raaijmakers21}
$12.28^{+0.50}_{-0.76}$ km \cite{Rutherford24}PP
$12.01^{+0.56}_{-0.75}$ km \cite{Rutherford24}CS
$R_{2.0}=12.33^{+0.70}_{-1.34}$ km \cite{Rutherford24}PP
$R_{2.0}=11.55^{+0.94}_{-1.09}$ km \cite{Rutherford24}CS

}

\section{Conclusions}
\label{sec:summary}

In this work we examined the combined effects of DM and MFs on DNSs properties.
We considered self-interacting asymmetric nonannihilating fermionic DM,
obeying the self-interaction cross section constraint
imposed by the observed interaction of galaxies in various galaxy clusters.
This left the DM particle mass as a free parameter.
Combined with the nucleonic QMC-RMF4 EOS
and density-dependent MFs of magnetar size,
we investigated how DM particle mass, DM mass fraction, and MF strength
influence key DNS properties such as maximum mass, mass-radius relations,
tidal deformability,
and the critical DM mass fraction needed for destabilization.

As is well known,
the DM-halo or -core character of a DNS is mainly determined
by a small (halo) or large (core) DM particle mass
(compared to the nucleon mass),
associated with possible increase or decrease of the
DNS maximum gravitational mass, respectively.

We found that the influence of the MF on these features is generally very small,
even for the strongest field values of magnetar size.
The magnetized EOS is softer
and consequently causes slightly smaller maximum masses and
less stability against collapse, for example,
for otherwise unchanged parameters.
But as the same effects are caused by a small variation of the DM fraction,
it will be a real challenge to extract this information from observation,
if such DNSs with large DM fractions exist.
We compared our results with data from PSR J0030+0451, PSR J0740+6620, and GW170817 to see which DM properties match their individual current constraints on mass, radius, and tidal deformability. We found that DNSs with a high fraction of DM are strongly limited by tidal deformability, while those with a lower DM fraction fit within observational data for a wide range of DM masses. A strong magnetic field slightly shifts these limits, mainly affecting the maximum possible NS mass.


\begin{acknowledgments}

B.~K.~acknowledges partial support from the
Department of Science and Technology, Government of India,
with grant no.~CRG/2021/000101. Part of this computation was performed using the CINECA cluster under the NEUMATT project.

\end{acknowledgments}

\section*{Data Availability}
No data were created or analyzed in this study.

\newcommand{\aap}{Astron. Astrophys.}
\newcommand{\apjl}{Astrophys. J. Lett.}
\newcommand{\araa}{Annu. Rev. Astron. Astrophys.}
\newcommand{\epj}{Eur. Phys. J.}
\newcommand{\jcap}{JCAP}
\newcommand{\jpg}{J. Phys. G}
\newcommand{\mnras}{Mon. Not. R. Astron. Soc.}
\newcommand{\npa}{Nucl. Phys. A}
\newcommand{\ptep}{Prog. Theor. Exp. Phys.}
\newcommand{\physrep}{Phys. Rep.}
\newcommand{\plb}{Phys. Lett. B}
\newcommand{\ssr}{Space Science Reviews}
\bibliographystyle{apsrev4-1}
\bibliography{adm}

\end{document}